\documentstyle[prl,aps,epsf,12pt]{revtex}
\begin{document}
\def\inseps#1#2{\def\epsfsize##1##2{#2##1}\centerline{\epsfbox{#1}}}
\newcommand {\beqa}{\begin{eqnarray}}
\newcommand {\eeqa}{\end{eqnarray}}
\newcommand {\n}{\nonumber \\}
\newcommand {\beq}{\begin{equation}}
\newcommand {\eeq}{\end{equation}}
\newcommand {\om}{\omega}
\newcommand {\s}{\sigma}
\newcommand {\Si}{\Sigma}
\newcommand {\de}{\delta}
\newcommand {\pa}{\partial}
\newcommand {\e}{\epsilon}
\newcommand {\Lm}{\Lambda}
\newcommand {\th}{\theta}
\newcommand {\ga}{\gamma}
\newcommand {\bt}{\bigtriangleup}
\newcommand {\bd}{\Diamond}
\newcommand  {\cl}{\clubsuit}
\newcommand  {\ov}{\overbrace}
\newcommand  {\un}{\underbrace}
\newcommand  {\nin}{\noindent}

\title{
Spontaneous Scaling Emergence in Generic Stochastic Systems }
\author{
Sorin Solomon and Moshe Levy  \thanks{ Email:sorin@vms.huji.ac.il, shiki@astro.huji.ac.il;
 http://shum.huji.ac.il/\~{} sorin }\\
Racah Institute of Physics,
Hebrew University, Jerusalem 91904, Israel\\
}
\date{21 August 1996}
\maketitle
\begin{abstract}
{\bf
We extend a generic class of systems which have previously been shown
to spontaneously develop scaling (power law) distributions of
their elementary degrees of freedom. 

While the previous systems were linear and exploded
exponentially for certain parameter ranges,
the new systems fulfill nonlinear time evolution equations
similar to the ones encountered in Spontaneous Symmetry Breaking (SSB)
dynamics and evolve spontaneously
towards "fixed trajectories" indexed by the 
average value of their  degrees of freedom (which corresponds to the
SSB order parameter).
The "fixed trajectories" dynamics evolves  on the edge between
explosion and collapse/extinction.

The systems present power laws with exponents which
in a wide range ($\alpha < -2.$) are universally determined 
by the ratio between the minimal and the average values
of the degrees of freedom.  
The time fluctuations are governed by Levy distributions of corresponding power.
For exponents $\alpha > -2$
there is no "thermodynamic limit" and the fluctuations are dominated
by a few, largest degrees of freedom which
leads to macroscopic fluctuations, chaos and bursts/intermitency.
}
\end {abstract}
\vfill
\eject


The wide-spread appearance of scaling laws in various natural systems
has attracted the attention of scientists for the last century
\cite{pareto}, \cite{zipf}, \cite{mandel}.
While for a measure zero set of systems
(statistical mechanics systems in equilibrium at critical
values of the parameters) scaling was thoroughly understood
\cite{kadanoff}, \cite{fisher}, \cite{wilson},
for most of the scaling systems in nature,
the origins of scaling remains a mystery \cite{gelmann}.
It was felt for the last decade that in many cases the emergence 
\cite{kauf} of scaling
has to do with complexity and self organization, i.e. some systems
have a tendency to organize spontaneously into critical states
\cite{bak}. Moreover, the identification of the relevant degrees of freedom
of a complex system was shown to be essential in understanding
its multiscale dynamics \cite{ANNREV} in terms of its
"irreducibly complex cores" \cite{nathan}.

In a recent publication \cite{IJMPC} it was shown that the appearance of
 the scaling  power  laws
is as generic in multiplicative stochastic systems as the Boltzmann law
is in additive stochastic systems.
By using the logarithm of the normalized degrees of freedom, we showed
that even non-stationary systems can be investigated by mapping them  into
additive-stochastic systems similar to usual
statistical-mechanics systems in thermal equilibrium.

In the present paper we extend our mechanism to
include a series of new features which makes it
applicable to a wider range of phenomena.

In particular we explain the natural
emergence of "fixed trajectories" for which,
(as opposed to the economics example in \cite{IJMPC})
the average mean value does not diverge in time.
  
These "fixed trajectories"
are analog to the "fixed points" in equilibrium statistical mechanics systems 
and naturally evolve  on the edge between exponential divergence and exponential decay. 

We show that on these "fixed trajectories", the systems
display power laws and Levy fluctuations distributions
which have wide universality basins with computable exponents.

For exponents $\alpha > -2$ of the  Levy distributions, one expects
the inexistence of a "thermodynamic limit" at large system sizes
and the
appearance of bursts and intermitency dominated by  the
fluctuations of the values of  a few  individual degrees of freedom
(low dimensional chaos).

We call our generic mechanism Spontaneous Scaling Emergence (SSE) as
it does not  require criticality or self-organization to achieve the scaling state.
The effective time evolution equations have  common features with the
equations of Spontaneous Symmetry Breaking. 

Before presenting the extensions let us discuss from the novel point
of view certain aspects of the systems introduced in \cite{IJMPC}. 
These systems are
characterized by the following stochastic time evolution equation :

\begin{equation} w_i (t+1) = \lambda w_i (t);
 \ \ \ \ \ \ \  i= 1,2, ..., N
\label{mult}
 \end{equation}

For definiteness one can think of the example of economic systems
\cite{IJMPC}, where $w_i (t)$
represents the wealth of investor $i$ at time $t$.
$\lambda$ is a random variable drawn from a generic distribution
$\Pi (\lambda,t) $.
It was imposed in \cite{IJMPC} that there
is some lower bound to the value of $w_i (t) $:

\begin{equation} 
w_i (t) \geq 
w_0\ {\bar w (t)}
\ \ \ \ \ \ \  \forall i,t
\label{bound}
\end{equation}

\noindent
where
\begin{equation} {\bar w (t)} = 1/N \sum_{i=1}^{n}w_i (t).
\label{sum}
 \end{equation}

\nin
This dynamics was shown to lead to a power law distribution for the various
$w$ values :

\begin{equation} \Phi(w, t) = C(t) w^{\alpha}
\label{power}
 \end{equation}

\nin
where $C(t)$ is a normalization factor and $\alpha = {-1-{1 \over 1-w_0}}$.
For  values of $ w_0$  small compared to 
$1/N$ (i.e. $-\log w_0 > \log N$) 
there are corrections  to (\ref{power}) originating in the contributions of large $w$ values.
The large $w$ corrections are relevant only for power laws 
with exponents $\alpha$ more positive than $-2$ and were discussed in \cite{IJMPCold}. 
For the sake of brevity we will not discuss them here again
in the new context.
One also expects a deviation from (\ref{power}) 
in the immediate vicinity of  the lower cut-off $w_0\ {\bar w(t)}$.

With these restrictions,
the striking property of (\ref{power}) is its independence on $\Pi (\lambda,t)$
except for the coefficient $C(t)$ which might depend on time in a quite
irregular fashion.

Since we are looking here for systems in which $C(t)$ flows naturally and
robustly to  finite  $C$ values, it is instructive to recall in some detail
how the robustness of (\ref{power}) arrised in \cite{IJMPC}. The crucial step was
to consider in place of the time-dependent distribution (\ref{power})
the distribution of the normalized variables

\begin{equation} \omega(t) = w_t / {\bar w(t)}
\label{normalized}
 \end{equation}

The distribution $ P(\omega , t)$ of the variables $\omega(t)$
is "pinned" at least against overall rescalings
$\omega  \longrightarrow K(t)\omega $ by the condition

\begin{equation} \int \omega P(\omega , t) d\omega = 1
\label{normalizationP}
 \end{equation}

\noindent
which follows trivially from the normalization (\ref{normalized}).
Equation  (\ref{mult}) becomes in terms of the new variables:

\begin{equation} \omega_i (t+1) = \mu \omega_i (t)
\label{multom}
 \end{equation}

\noindent
where the distribution $\Pi' (\mu )$ of the random variable
$\mu$ is related to the distribution $\Pi (\lambda )$ of $\lambda$
through the relation

\begin{equation} \Pi'(\mu,t) = \Pi(  {{\bar \omega{(t)}} \over {\bar \omega{(t+1)}}}\mu, t)
{{\bar \omega{(t+1)}} \over {\bar \omega{(t)}}}
\label{pi`}
 \end{equation}

\noindent
This overall rescaling in the argument of $\Pi $ turned out to be enough to
eliminate the dependence of $C$ on $t$ in (\ref{power}).
This means that the use of $\omega's$ in place of $w$'s rescales in a very particular
way the argument of $\Pi$.
More precisely $\Pi'(\mu,t)$ is fine-tuned by the transformation
(\ref{normalized}) such that
$P(\omega , t+1) = P(\omega , t)$, i.e. (eq. 9 in Ref [12])

\begin{equation} \int \Pi'(\mu,t) \mu^{1+{1 / ( 1-\omega_0)}}d\mu = 1
\label{normalizationPi'}
 \end{equation}

This steady state situation is obviously reached by having the
distribution of the $\mu's$ which appear in (\ref{multom})
finely tuned to equilibrate the values of $\mu$ larger than 1
(which lead to the increase of the $\omega$'s)
with the smaller $\mu$ values (which decrease the $\omega$'s).

One sees now that one could in principle allow the shape of the distribution
$\Pi (\lambda,t)$ to depend on $w$ as long as an overall
rescaling of its parameter
$\lambda \longrightarrow {{\bar \omega{(t)}} \over {\bar \omega{(t+1)}} }\lambda $
insures that the distribution stability condition (\ref{normalizationPi'}) 
holds for each $\omega$ i.e.
$ \int \Pi'(\mu,t,\omega ) \mu^{1+{1 / ( 1-\omega_0)}}d\mu = 1
\ \ \ \forall \omega ,t$.
In terms of the stock market this means that the power law is
stable even if the spread (width) of the gain-loss
distribution depends on the wealth of the investor
(or varies in time). All that is
required is that the average gain expectancies of all
the investors are equal (except for the poorest, which can not lose
more than allowed by the poverty bound $w_0$).

This is quite a large degree of universality even though unlike
critical phenomena of statistical
mechanics the scaling exponents depend here on a continuous
parameter.

Now it is clear what is needed in order to get rid of the time dependence
of $C(t)$ in (\ref{power}):
one needs to find a mechanism which brings
$\Pi(\lambda)$ to be finely balanced
between $\lambda > 1$ and $\lambda < 1$ values, such as to fulfill
equation (\ref{normalizationPi'}). Such a mechanism will prevent the exponential
explosion or collapse of (\ref{mult}). 
In the stock market example this
mechanism is expressed by the fact that 
as long as the total real wealth of the community is unchanged,
overall gain factors across the entire population
are irrelevant (they express just inflation in the instruments of payment). 
It will turn out
that this is a degenerate example of a wider generic
phenomenon.

In order to see how our new models exploit this idea,
consider a system governed by the stochastic evolution rule :

\begin{equation} \omega_i (t+1) - \omega_i (t) = \kappa
(\omega_i (t) + \omega_0)  ;
 \ \ \ \ \ \ \  i= 1,2, ..., N
\label{mult2}
 \end{equation}

\begin{equation} \omega_i (t) \geq 0
\ \ \ \ \ \ \  \forall i,t
\label{bound2}
 \end{equation}

\nin
where $\kappa$ is a random variable drawn from a distribution
$\Upsilon(\kappa, {\bar \omega (t)})$.
We use different symbols for
the analogues of $\lambda$ and $\Pi$ of (\ref{mult}) in order to
avoid possible confusion. Moreover note that
in the new models there is
no normalization condition on the elementary variables $\omega_i (t)$ and therefore
their average $\bar \omega (t)$ is a variable. 
In fact, $\Upsilon(\kappa, {\bar \omega (t)})$
depends on  ${\bar \omega (t)}$ and is assumed that for ${\bar \omega (t)}= 0$
the distribution
$\Upsilon(\kappa,0)$ favors the appearance of strongly positive values of $\kappa$
while for ${\bar \omega (t)}$ at a larger scale $W$,
the distribution
$\Upsilon(\kappa, W)$  favors negative values of $\kappa$.
According to (\ref{mult2}), this means that for small values of ${\bar \omega (t)}$
the $\omega's$ will have the tendency to increase while for
 ${\bar \omega (t)} \approx W$ they will have a tendency  to decrease.
This means that there exists a value ${\bar \omega_{eq}}$
(determined by a condition of the type (\ref{normalizationPi'})
towards which ${\bar \omega (t)}$  will always flow, by the virtue of
(\ref{mult2}).(We defer to further publications
the discussion of non-monotonic $\Upsilon$'s which lead to
multiple solutions ${\bar \omega_{eq}}$ and to possible  branchings/bifurcations).

Once  ${\bar \omega_{eq}}$ is reached,  eq. (\ref{mult2}) becomes of the
type (\ref{multom}) (for the variable $\omega_i (t) + \omega_0$).
Independently of the details of $\Upsilon(\kappa, {\bar \omega (t)})$
one then has :

\begin{equation} P(\omega) = C (\omega + \omega_0)^
{-1-{1 / ( 1-\omega_0 / {\bar \omega_{eq}} )}}
\label{powereq}
 \end{equation}

\nin
with constant $C$.
This is a stationary power law, similar in form to (\ref{power}), but with
no time dependence !
According to the analysis in \cite{IJMPC},
the time fluctuations of the system are then characterized
by a random walk with steps of variable stochastic length
distributed according to (\ref{powereq}).
This means , according to the generalized central limit
theorem, that the time fluctuations of various quantities
are given by a Levy distribution \cite{levy} of corresponding power.

In the case in which the Levy distribution is wider (and 
the the power law exponent $ \alpha $ is $> -2.$ due to the effect
of large $w_i$ values \cite{IJMPCold}),
large macroscopic fluctuations are expected due to the variations
of the large $w_i$'s.
In particular, they might bring a system macroscopically
away from the fixed point ${\bar \omega_{eq}}$.
This leads to a burst of non-equilibrium activity until
enough correcting (relaxation) steps take place. 
In the case that it exists more then one ${\bar \omega_{eq}}$ value,
these macroscopic fluctuations may induce jumps between the various "equilibrium"
trajectories.
The fact that for $ \alpha > -2 $  Levy distributions
only the largest elements are relevant to the dynamics 
predicts the absence of a thermodynamic limit
as the number of elementary degrees of freedom is taken to infinity.
Such a phenomenon might explain some of the simulation results in
\cite{nathanp},\cite{Hell}.
We propose to study
these effects numerically and compare with the theoretical and experimental
universal properties of chaotic systems \cite{Fiegenbaum}.

The assumption in (\ref{mult2})  that the distribution
$\Upsilon(\kappa, {\bar \omega (t)})$ depends on ${\bar \omega (t)}$
might seem arbitrary, but in fact it arises naturally in
quite familiar contexts. Consider a generic complex system
in which some cooperative "short-range" underlying dynamics
insures that certain subsystems act coherently and can be represented
by a single variable $\omega_i (t)$.
One can think of such subsystems as families, clans,
populations within an ecology, etc.
Assume that isolated from each other these systems act according to
(\ref{mult}). Suppose however that there is an "infinite-range"
competitive interaction between the $\omega$'s (clans
competing for land, populations for food, etc.).
The resulting dynamics may then be of the form

\begin{equation}
\omega_i(t+1) - \omega_i(t)  =  \kappa \omega_i(t) + \omega_0{\bar \omega (t)} -
{a \over 2}{\bar \omega (t)} \omega_i (t) + \ etc. \ ,
\label{example}
 \end{equation}

where $etc.$ stands for possible additional inhomogeneous  terms.
\nin
One sees that for small values of $\bar\omega$ the stochastic
equation (\ref{example}) drives the $\omega$'s to higher values
while for large values of $\bar\omega$ the $\omega$'s are driven
to smaller values and one is in the case (\ref{mult2}).

Note that equation (\ref{example}) is very similar to the familiar equation
governing the emergence of Spontaneous Symmetry Breaking (SSB):

\begin{equation}
{\dot M}  =  \kappa M - {a \over 2}M^2 + \  etc.
\label{magnet}
 \end{equation}

\nin
from which it differs only by the fact that it mixes the
"short-range" variable $\omega_i (t)$ with the "infinite-range"
variable $ {\bar \omega (t)} $. 

This suggests the extension of our models for 
"vectorial" $\omega$ variables  which take values in spaces with
various symmetries. 

In terms of the SSB vocabulary, the power of the distribution (\ref{powereq}) is
governed by the ratio between the scale of the global variable
$ {\bar \omega (t)} $ and the scale of the local variable
$\omega_0$. As seen above, both of them may be dynamically fixed by
the interplay of various non-homogeneous terms in (\ref{example}).
Note that the emergence of power laws and hierarchies follows then
naturally even if the ratio of the 2 scales is  of order $1$
(e.g. $0.2 - 0.3$ in the economics application \cite{IJMPC}).

In further studies one can consider the effects of higher powers
in (\ref{example}) and relax the
"infinite-range" assumption. In fact, one can study multiscale
hierarchies of ranges, and consider the domains of almost constant
$\omega$'s as the elementary degrees of freedom $\Omega$ of a coarser system.
For instance, each family within a clan acts as a unit but
competes with the other families for the clans resources.
Yet at a larger scale, the clan acts as a self-reinforcing unit
in competition with other clans over the country resources, and
so on. Examples in diluted finite connectivity spin glasses
\cite{nathan} and  Weiss domains in magnetic systems may
be of relevance.

This multiscale cooperation-competition interplay might be
particularly relevant for the study of the speciation process
in various ecological and territorial conditions.
It also raises some suggestions on the ways to implement the
reductionist program as one passes from one scale to another.
The scaling laws and their relevant degrees of freedom might give
some objective criterion for identifying the passage from one
level of description to another. In the spirit of \cite{ANNREV}
and \cite{nathan} one can "turn the tables" and use the power
laws in order to identify from (\ref{powereq}) the size and
identity of the relevant effective degrees of freedom of the
system: the "irreducibly complex cores" in the terminology of
\cite{nathan}. In turn, the dynamics of $\bar\omega$ would
indicate the "multiscale-complex" dynamics of the hierarchy of
macros in the terminology of \cite{ANNREV}.

Let us reiterate in conclusion the main points discussed in the present paper.
\begin{enumerate}
\item We have shown that systems with local self-catalizing  multiplicative  interactions
and  competitive global interactions (\ref{mult2})
(\ref{example}), may generically flow into "fixed trajectories" 
at the edge between explosion and collapse.
\item The various "fixed trajectories" are indexed by the values of the
average degrees of freedom which emerge from a nonlinear
time evolution equation similar in form to the one governing SSB. 
\item In those equations, both local and global variables appear
and their scales ($\omega_0$ and ${\bar \omega}$) may be fixed by the 
generic nonlinear dynamics.
\item The degrees of freedom are distributed according  power laws
(\ref{powereq})
whose exponents are determined universally (for a wide range) only by the
ratio of the 2 scales ${{\omega_0} / {\bar \omega}}$.
\item The fluctuations around the "fixed trajectories"
are governed by  Levy distributions of corresponding (universal) exponents.
\item
In the cases in which the Levy distribution has exponent
$ \alpha > -2 $
the system does not have a thermodynamic limit and presents
bursts and intermitency produced by the few degrees of freedom with the
largest values (chaos \cite{Fiegenbaum}). 
\end{enumerate}

We are now facing the enormous task of applying the present
framework to particular complex systems beyond the
economic example presented in \cite{IJMPC}.
The most urgent objective is to relate for each system the
complex dynamics to the appropriate   $\omega$'s, identify
$\omega_0$ and relate it by (\ref{powereq}) to the actual power
law measured experimentally.

The values of $\omega_0$  may be determined by different
mechanisms in various systems. Examples for such mechanisms
are the extinction of species with too few members, subsidies to
the poor, minimal size of population needed to found a city,
discretization, additive noise, pumping of energy,  etc.

\eject

\noindent
Note Added in Proof
\noindent

A detailed discussion of some generalizations of the model
introduced in \cite{IJMPC} can be found in
"Convergent Multiplicative Processes Repelled from Zero"
by D. Sornette and R. Cont. We thank D. Sornette for
very instructive correspondence. We thank D. Stauffer for
advice and encouragement since the beginning of our research
on the present subject.


\begin{thebibliography}{99}

\bibitem{pareto}
{Pareto, V. {\em Cours d'Economique Politique}, Vol 2, (1897).}

\bibitem{zipf}
{ Zipf, G. K. {\it Human Behavior and the Principle of Least Effort}
(Addison Wesley, Cambridge MA, 1949).}


\bibitem{mandel}
{Mandelbrot, B. B. {\it Comptes Randus} {\bf 232,}  (1951) 1638-1640.}

\bibitem{kadanoff}
{{Kadanoff, L. } {\em Physics}{\bf 2,} (1966) 263.}


\bibitem{fisher}
{Fisher, M. E. {\em Reviews of Modern Physics}{\bf 46,} (1974) 597.}


\bibitem{wilson}
{Wilson, K. G. {\em Reviews of Modern Physics}{\bf 47,} (1975) 773.}

\bibitem{gelmann}
{Gell-Mann, M. {\it The Quark and the Jaguar}, 92-106
(Little Brown and Co., London, 1994).}

\bibitem{kauf}
{Stuart A Kauffman, {\em At home in the universe}, Oxford
University Press, New York 1995,}

\bibitem{bak}
{Bak, P., Tang, C. \& Wiesenfeld, K. {\em Phys. Rev. Lett.} {\bf 59,} (1987) 381.}

\bibitem{ANNREV}
{S. Solomon,
{\em The Microscopic Representation of Complex Systems},
in Annual Rev. of Comput. Phys. II,
ed. D. Stauffer, World Scientific, Singapore 1995.}

\bibitem{nathan}
{Persky, N and Solomon, S.
{\em Phys. Rev. E.} (1996) to appear.}


\bibitem{IJMPC}
{ Levy, M.,  and Solomon, S.
 {\em Intn. J. Mod. Phys. C} {\bf 7,} No. 4 (1996) 595.}

\bibitem{IJMPCold}
{ Levy, M.,  and Solomon, S.
 {\em Intn. J. Mod. Phys. C} {\bf 7,} No. 1 (1996) 65.}

\bibitem{nathanp}
{M. Levy, N. Persky and S. Solomon,
{\em Int. J. High Speed Computing} to appear (1966).}

\bibitem{Hell}
{T. Hellthaler,
{\em Int. J. Mod. Phys. C} {\bf 6}  (1995) 845.}

\bibitem{Fiegenbaum}
{M. J. Fiegenbaum
{\em J. Stat. Phys.} {\bf 19} 25 (1978) 669.}

{E. N. Lorentz
{\em J. Atmos. Sci.} {\bf 20} (1963) 130.}

{R. M. May
{\em Nature} {\bf 261}  (1976) 459 .}



 
\bibitem{levy}
{P. Levy,
{\em Theorie de l'Addition des Variables Aleatoires}
(Gauthier-Villiers, Paris, 1937).}



\end{thebibliography}
\end{document}